\documentclass{article}
\usepackage{spconf,amsmath,graphicx,hyperref}
\usepackage[most]{tcolorbox}
\title{RealClass: Real Educational Speech in Virtual Acoustic Classroom Environments}
\title{RealClass: Real Educational Speech in Virtual Acoustic Classroom Environments for Automatic Speech Recognition}
\title{RealClass: A Framework for Classroom Speech Simulation with Public Datasets and Game Engines}
\name{Ahmed Adel Attia\textsuperscript{1},  Jing Liu\textsuperscript{2}, Carol Espy Wilson\textsuperscript{1}\thanks{This work is supported by the Grand Challenge Award at the University of Maryland and through the generous support of the Bill and Melinda Gates Foundation}}
\address{University of Maryland\\
\textsuperscript{1}Department of Electrical and Computer Engineering, \textsuperscript{2}College of Education\\
aadel@umd.edu, jliu28@umd.edu, espy@umd.edu}
\begin{document}
%\ninept
%
\maketitle
\begin{abstract}
The scarcity of large-scale classroom speech data has hindered the development of AI-driven speech models for education. Classroom datasets remain limited and not publicly available, and the absence of dedicated classroom noise or Room Impulse Response (RIR) corpora prevents the use of standard data augmentation techniques. 

In this paper, we introduce a scalable methodology for synthesizing classroom noise and RIRs using game engines, a versatile framework that can extend to other domains beyond the classroom. Building on this methodology, we present RealClass, a dataset that combines a synthesized classroom noise corpus with a classroom speech dataset compiled from publicly available corpora. The speech data pairs a children’s speech corpus with instructional speech extracted from YouTube videos to approximate real classroom interactions in clean conditions. Experiments on clean and noisy speech show that RealClass closely approximates real classroom speech, making it a valuable asset in the absence of abundant real classroom speech. 
\end{abstract}
\begin{keywords}
Classroom speech, Educational speech data, Noise simulation, Room impulse response (RIR), Automatic speech recognition (ASR), Robust speech processing\end{keywords}
\vspace{-10pt}

\section{Introduction}
\label{sec:intro}

The performance of AI models is defined by its training data, and the model's ability to learn from this data. Speech models are no exception. For instance, Whisper \cite{radford2023robust} was able to achieve state-of-the-art performance in the Automatic Speech Recognition (ASR) task by training on more than half a million hours of transcribed speech scrapped from the internet. 

Naturally, this scalability does not apply to low-resource settings. For the same ASR task, low-resource languages lag behind English and other high-resource languages precisely due to the lack of high-quality labeled data for these languages \cite{san2024predicting, nowakowski2023adapting}. Beyond linguistic variability, even some English language tasks suffer from data scarcity. Challenging acoustic conditions like high background noise and multi-speaker environments usually require more training data for robust performance. However, for some domains, like classroom speech and children's speech, there is a severe data scarcity, since children are considered a protected class and speech is considered Personally Identifiable Information (PII) \cite{coppa2024}.  As a consequence, releasing public datasets including children's speech is complicated.

Prior research into developing classroom and children ASR tools identified data scarcity as a major hurdle to improving performance \cite{attia2024cpt, southwell2024automatic}. While there are some children ASR datasets that are public \cite{pradhan2023my, shobaki2000ogi},  most classroom corpora are not. As a result, different researchers working on the same problem often work with different datasets. Getting access to protected classroom speech corpora is a complicated task, so each research group tends to collect their own data, but cannot publish it \cite{attia2024cpt, southwell2022challenges}. This lack of dataset exchange negatively affects the progress of research and reproducibility of research.

Some prior research has investigated data augmentation methods that adjust the acoustic properties of adult speech to match that of children, such as pitch perturbation \cite{fan2024benchmarking}. While these methods are helpful in mimicking the acoustic properties of children speech, they do not capture their unique linguistic properties \cite{attia2023kid}. They also do not address the main defining challenges of classroom speech. Primarily, classroom speech suffers from a high level of children's babble noise, which is the type of noise resulting from multiple background speakers overlayed with speech from the target speaker. While adult babble noise corpora exist \cite{snyder2015musan, font2013freesound}, no such corpus exists for children's babble noise. 

In this paper, we present our novel dataset curation framework that aims to solve both issues: the scarcity of classroom data and the lack of children's babble noise corpora. We create a clean classroom noise corpus by combining the My Science Tutor (MyST) corpus with instructional adult speech from lectures from MIT OpenCourseWare (OCW) and Khan Academy. These channels' videos are licensed under Creative Commons BY-NC-SA \cite{cc-by-nc-sa}, which permits non-commercial use, adaptation, and redistribution with proper attribution. In addition, since classroom speech is noisy by nature, we synthesize classroom noise, primarily children's babble, through the Unity Game Engine. The Unity Game Engine provides high-fidelity audio simulation in 3D virtual space, which accurately simulates the acoustic characteristics of the classroom environment, as well as the spatial characteristics of different audio sources. We use the same software to calculate Room Impulse Responses (RIRs) from the virtual classrooms to create the first classroom-specific RIR bank.

We call our dataset \textbf{RealClass}\footnote{In a previous preprint, we called a predecessor to this dataset SimClass. We opted to rename the dataset to emphasize that the speech in this dataset is not generative, and to avoid confusion with other techniques that share the same name.}. We plan to make RealClass and the tools developed to create it publicly available at no cost to researchers. RealClass is the largest and only public classroom speech dataset. In addition, the existence of a clean and noisy version of the dataset, which is not possible with real classroom recordings with naturally occurring noise, opens the door to many tasks that were not possible, such as speech enhancement. The reason for that is that, unlike real noisy data, RealClass provides clean-noisy pairs for the same audio, required to train speech enhancement model. RealClass mainly simulates elementary school STEM classes as a result of its constituting data, but our methodology outlined in this paper paves the way for simulating other kinds of classrooms given different kinds of data.

Our work is the first to demonstrate that classroom noise and RIRs can be synthesized at scale, validated through ASR experiments, and released publicly, filling a gap that neither augmentation nor private corpora can address.
\vspace{-10pt}
\section{Datasets}
\label{sec: data}
\subsection{Datasets Used To Create RealClass}
\subsubsection{My Science Tutor (MyST)}
The MyST corpus is the largest publicly available children's speech corpus. It consists of 393 hours of conversational children's speech, recorded from virtual tutoring sessions in physics, geography, biology. The corpus spans 1,371 third, fourth, and fifth-grade students. Around 210 hours of the corpus were accurately transcribed after filtering out weak, inaccurate transcriptions following the method outlined our prior work \cite{attia2023kid}. We utilize the transcribed portions of the children's speech for our clean speech base. We use the speech from the untranscribed portion for noise simulation.% We have verified that there is no overlap in speakers between the transcribed and untranscribed portions, to avoid having the target speech from the primary child speaker also be present in the background.
\subsubsection{Khan Academy}
Khan Academy is a popular YouTube channel hosting thousands of educational videos targeted at different school-age grades. Although Khan Academy videos provide a good match in both topic and target audience, as there is a variety of elementary STEM playlists, they mainly have the same primary speaker, which can limit the diversity of the data. However, to benefit from the agreement on the subject matter and topics, we include videos from this channel in our dataset. 

We chose 7 playlists discussing topics in second and third-grade mathematics. The total duration of the videos in these playlists is about 7.5 hours, and they include high-quality audio accompanied by human transcripts.

\subsubsection{MIT OpenCourseWare (OCW)}
MIT OCW \cite{MITOCW_YouTube} is an online repository of recorded lectures from 2,500 MIT courses. Although the classes in OCW are college-level and graduate-level classes, they still provide a useful resource for elementary school STEM classes, as the acoustic properties of instructional speech are evident in the videos. However, to help bridge the gap between the linguistic components of college-level courses and STEM classes, we picked videos from 10 courses in calculus and linear algebra, totaling around 174 hours. 

These videos include high-quality, clear audio from a variety of lecturers with different accents, accompanied by high-quality human transcriptions. We will provide a full list of videos used on our GitHub repository in the camera-ready manuscript.

\subsection{Classroom Datasets Used For Testing: NCTE Dataset}
% \subsubsection{NCTE Dataset}
The NCTE dataset \cite{demszky2022ncte} is a collection of video and audio recordings of 2128 4th and 5th-grade mathematics classrooms. Out of these recordings, only 17 are transcribed, highlighting the difficulty of large-scale human transcription. These recordings amount to 12.8 hours of speech. We set aside 4 classroom recordings for a 2.9-hour test/validation set, and use the rest for training a benchmark model that represents the performance using existing classroom data.
% \subsubsection{M-Powering Teachers (MPT) Dataset}

% The MPT dataset is a collection of 6 classroom recordings that took place in 2023. These classrooms were distributed between California, Ohio, and Washington D.C., and range from the 5th to the 8th grade. The total duration of this dataset is 3 hours. We reserve this dataset entirely for testing.
\vspace{-10pt}

\section{Methodology}
\label{sec: method}
\subsection{Creating The Clean Classroom Speech Base}
To create the clean version of RealClass, we combine individual tracks from the adult corpora with tracks from the MyST dataset. To make the resultant tracks more semantically robust, we perform semantic matching between the adult and child corpora. Specifically, we encode the transcripts from both datasets with the SentenceTransformer \cite{reimers-2019-sentence-bert} model\footnote{\url{https://huggingface.co/sentence-transformers/all-mpnet-base-v2}}, and normalize the embeddings for similarity search. A FAISS\cite{douze2024faiss} index greedily matches children-adult utterance pairs based on semantic similarity. This process produces semantically consistent clean speech pairs that approximate realistic classroom interactions. Below is an example of a semantically matched transcription:
\vspace{-5pt}
\begin{tcolorbox}[colback=gray!10, colframe=gray!50, arc=2mm, boxrule=0.5pt, left=0pt, right=0pt, top=0pt, bottom=0pt]
% \scriptsize
    \textbf{Child:} It flows to the positive end to the negative end.\\
    \textbf{Adult:} What flows in the opposite direction?
\end{tcolorbox}

To create variety, for each combination, the file can either start with child speech followed by adult speech, or vice versa. In either case, we allow for a random overlap between 0.5 seconds and 1 second for 20\% of the files, simulating cut-off during turn-taking in normal conversations. 

The total duration of the RealClass dataset is 391 hours, making it the largest classroom speech corpus and the only publicly available one. As for the train/test/validation partition, we partitioned each constituent dataset before the combination. We follow the splits created by the authors of the MyST dataset, which ensures that no speaker exists in two splits. For the YouTube datasets, we partitioned them by channels to create a roughly 80/10/10 partition across both OCW and Khan Academy. As a result, we ended up with 313 hours of training data, 37 hours of development data, and 41 hours of test data.

\vspace{-10pt}
\subsection{Simulating Classroom Noises in Unity Game Engine}

To create our classroom babble noise, we use the Unity Game Engine. While game engines are typically associated with graphics, modern frameworks like Unity also support advanced acoustic simulations through plug-ins such as Steam Audio \cite{SteamAudio}. By leveraging the environment’s geometry, Steam Audio models effects such as diffraction, occlusion, and reverberation, and supports “acoustic materials” that define surface properties, such as absorption, transmission, scattering, for objects in the scene \cite{nieminen2021unity}.

We design a classroom environment by acoustically modeling all objects and surfaces, including chairs, desks, doors, windows, ceilings, and carpets. Within the classroom, we place 25 spatially directive audio sources playing tracks from the non-transcribed portion of the MyST Tutor dataset, representing 25 students. Each source has its own orientation and emission pattern, creating realistic overlapping speech that captures the character of children’s babble. To add further realism, we include randomly placed chair noises and ambient playground sounds, sourced from YouTube, at random intervals. Noise is captured using a moving virtual microphone (“audio listener”) to record from different angles, resulting in 50 hours of physically simulated classroom noise.
\vspace{-10pt}

\subsection{Measuring RIR from Virtual Environments}
An RIR characterizes how an acoustic environment transforms sound. Convolving an RIR with a clean audio signal produces a realistic rendering of how the signal would be perceived within that space. Building on this principle, we construct, to the best of our knowledge, the first dataset of classroom RIRs using a Unity-based simulation framework. This approach offers key advantages over directly re-recording audio in virtual environments: convolution is highly efficient, enabling hundreds of hours of audio to be simulated within minutes, and the resulting RIRs are easily shareable, allowing flexible reuse across diverse signals.

We follow the Exponential Sine Sweep (ESS) technique \cite{farina2000simultaneous} to calculate RIRs. The system is excited with a sinusoidal signal whose frequency increases exponentially across the audible range (20 Hz–20 kHz). The response is recorded and then deconvolved with the inverse of the excitation signal to obtain the RIR.

To construct a diverse RIR bank, we simulate eight distinct classroom environments. In each room, five source positions are selected (center and corners), and for each source, we record responses at the remaining positions. This source–receiver pairing yields 20 unique RIRs per room (160 in total), capturing variability in both sound emission and listening position.

\vspace{-10pt}

\section{Validation}
% In this section, we validate our dataset through experiments on the classroom ASR benchmark. We chose automatic speech recognition as the primary evaluation task because it is one of the most active areas of research in classroom speech. At the same time, our dataset is broadly applicable to other tasks such as speaker identification and diarization, and it enables new directions that are not possible with real classroom data, including speech enhancement.% In real recordings, there is no noise-free reference, making enhancement models difficult to train, whereas our dataset provides paired clean and noisy conditions by design.

 In this section, we validate our dataset through experiments on the classroom ASR benchmark. We use the Fairseq implementation of Wav2Vec2.0. Unless otherwise noted, models are initialized from the Robust-Wav2vec2.0\cite{hsu2021robust} checkpoint. We also use our Wav2vec2.0 model, specifically pre-trained for classroom speech \cite{attia2024cpt} through Continued Pre-training (CPT) on unlabeled classroom speech. We reserve the CPT model for training our final models that are compared to training on real classroom data. %For the baseline trained solely on classroom speech, we use our Wav2vec2.0 model, specifically pre-trained for classroom speech \cite{attia2024cpt} through Continued Pre-training (CPT) on unlabeled classroom speech. Even though this model always provides better results when finetuned with any dataset, we use the vanilla Robust-Wav2vec2.0 for the majority of our experiments to isolate the effect of CPT from our contributions.
\vspace{-10pt}
 
\subsection{Comparison with Non-Classroom Baselines}
To contextualize the performance of our proposed RealClass dataset, we compare against Librispeech, a non-classroom baseline that is commonly used in ASR research, as well as the individual components of RealClass. In addition, we evaluate simple data combinations of these components, as well as our proposed semantic pairing method that aligns adult and child speech into synthetic dialogues. Results are in Table \ref{tab:external_baselines}
\begin{table}[h]
\centering
\vspace{-10pt}

\caption{WER (\%) on classroom test data for models trained on non-classroom corpora. \textit{"Online Lectures''} refers to data from OCW and Khan Academy YouTube channels. RealClass outperforms all alternatives, showing the value of synthetic classroom simulation.} \vspace{1pt}
\label{tab:external_baselines}
% \resizebox{\columnwidth}{!}
%   {
\begin{tabular}{lc}
\hline

\textbf{Training Data} & \textbf{WER (\%)} \\
\hline
LibriSpeech & 40.64 \\
Online Lectures & 38.33 \\
MyST & 45.82 \\\hline
Online Lectures + MyST & 37.55 \\
\textbf{RealClass} & \textbf{35.52} \\
\hline
\end{tabular}
% }
\end{table}

 As expected, training on LibriSpeech yields the worst performance on classroom speech, with a WER of roughly 40\%. This result highlights the severe domain mismatch: although LibriSpeech is a large and clean corpus, it lacks the overlapping speech, noisy conditions, and mixed child–adult speech distributions that characterize real classrooms.

Training only on instructional lecture data from OCW and Khan Academy (referred to as \textit{"Online Lectures'' } in Table \ref{tab:external_baselines}) provides some improvement over LibriSpeech. This is likely due to a closer match in speaking style and domain vocabulary, as these lectures often resemble teacher discourse in classrooms. Although almost entirely dominated by teacher speech, incidental multi-speaker events (e.g., students asking questions) and the room acoustics present in the recordings introduce variability that overlaps with classroom conditions. 

Training exclusively on MyST, a corpus composed entirely of children’s speech, results in worse performance than LibriSpeech. This outcome is intuitive: although children’s speech is part of the classroom domain, MyST consists of single-speaker tutoring sessions and lacks the adult-dominated discourse that characterizes most classroom speech. Thus, training on children’s speech alone yields poor generalization to classroom test data. However, by simply concatenating instructional speech with MyST, performance imporoves over either dataset alone, as well as over LibriSpeech. This result suggests that combining adult instructional data with children’s speech provides a more balanced proxy for classroom conditions. Taking it a step further, our proposed RealClass methodology further improves performance by semantically pairing utterances from adult and child datasets into synthetic dialogues, with randomized overlaps to mimic interruptions. This curation process not only balances the linguistic roles of adults and children but also more closely approximates the turn-taking dynamics of real classrooms. RealClass serves as a strong baseline for the subsequent addition of simulated room acoustics and noise, validating our approach to dataset construction.

% % Table 1: Individual Ablations
% \begin{table}[h]
% \centering
% \caption{ASR performance (WER \%) for individual ablations of RealClass. 
% $\Delta$ WER (abs.) is the absolute WER reduction compared to training on RealClass alone.}
% \label{tab:realclass_individual}
%  \resizebox{\columnwidth}{!}
%   {
% \begin{tabular}{lcc}
% \hline
% \textbf{System} & \textbf{WER (\%)} & \textbf{$\Delta$ WER (abs.)} \\
% \hline
% RealClass & 35.52 & -- \\
% RealClass + Room Acoustics (RIR) & 33.52 & 2.00 \\
% RealClass + Children Babble Noise & 32.51 & 3.01 \\
% \hline
% \textit{Baseline} & \textit{21.12} & \textit{14.40} \\
% \hline
% \end{tabular}}
% \end{table}
\vspace{-10pt}

\subsection{Ablation Studies}
% Table 2: Cumulative System
This section examines both the individual contributions of synthetic realism (RIR, noise) and the cumulative contributions when combining synthetic and real classroom data. Results are shown in Table \ref{tab:realclass_cumulative}.

\begin{table}[h]
\centering
\vspace{-6pt}
\caption{Ablation study of RealClass with synthetic realism and continued pretraining. 
$\Delta$WER (abs.) is the absolute WER reduction compared to RealClass alone. The top block reports independent ablations of RealClass with RIR, additive noise, or both, while the bottom block reports cumulative improvements when further adding real classroom data (NCTE) to the training data and using the CPT Wav2vec2.0 model from \cite{attia2024cpt} } \vspace{1pt}
\label{tab:realclass_cumulative}
 \resizebox{\columnwidth}{!}
  {
\begin{tabular}{lcc}
\hline
\textbf{System} & \textbf{WER (\%)} & \textbf{$\Delta$ WER (abs.)} \\
\hline
RealClass & 35.52 & -- \\
RealClass + Room Acoustics (RIR) & 33.52 & 2.00 \\
RealClass + Children Babble Noise & 32.51 & 3.01 \\
RealClass + RIR + Noise & 32.76 & 2.76 \\
\hline
\multicolumn{3}{l}{\textit{Cumulative additions:}} \\
+ CPT& 26.24 &	9.28\\

+ Additive Real Classroom Data (NCTE) & \textbf{19.98} & \textbf{15.53} \\
\hline
\textit{Baseline: Only Real Classroom Data (NCTE)} & \textit{21.12} & \textit{14.40} \\
\hline
\end{tabular}}

\vspace{2pt}
\end{table}

% From the previous section, we know that clean RealClass serves as a strong starting point: it outperforms non-classroom baselines and provides a better approximation of classroom conditions than training on individual external datasets. This validates our design choice of semantically pairing adult and child speech into clean synthetic dialogues.

As shown in Table 2, convolving RealClass with simulated room impulse responses (RIRs) reduces WER by about 2\% absolute, indicating that the addition of classroom-like acoustics makes the synthetic data more consistent with the test conditions. Similarly, adding simulated classroom babble noise improves WER by about 3\%, highlighting the effectiveness of modeling the noisy conditions typical of real classrooms. These results suggest that both acoustics and noise individually push the synthetic data closer to real classroom distributions. However, when RIR and noise are combined, performance slightly degrades. We attribute this to an artifact of the current pipeline: the clean speech is first convolved with an RIR, while the noise track already contains reverberation from the Unity simulator. Mixing the two may produce mismatched acoustics, which may affect the performance. We treat this as a current limitation of the dataset and leave matched-acoustics rendering for future work.

When switching from the robust Wav2Vec2.0 model to our continued pretraining (CPT) model, which has been adapted to classroom data in prior work, performance improves substantially (from 32.76\% to 26.24\% WER). This confirms earlier findings that CPT is the most effective initialization for classroom ASR and underscores why we kept earlier ablations fixed to the robust model, to isolate the effect of CPT. Comparing against the NCTE baseline, RealClass+RIR+Noise+CPT achieves 26.24\% WER, while NCTE training achieves 21.12\% WER. RealClass provides a strong approximation when such data is unavailable.

Finally, combining RealClass with NCTE yields the best performance of 19.98\% WER, surpassing the use of NCTE alone. This demonstrates a second use case for our dataset: not only can it serve as a standalone substitute when real classroom data is scarce, but it can also act as a complementary resource to real data, further enhancing performance when the two are combined.
\vspace{-10pt}

\section{Conclusion and Broad Impact}

In this paper, we introduced RealClass, a simulated classroom dataset that integrates clean adult and child speech with room impulse responses and classroom noise. Our experiments demonstrate that these components, when combined, provide a strong approximation to real classroom speech.

The results highlight two key use cases. (1) In the absence of real classroom data, which is a common challenge due to privacy constraints and collection costs, RealClass serves as a close analog, achieving performance comparable to models trained on real data. (2) In scenarios where limited classroom data is available, combining RealClass with real classroom recordings yields further improvements, surpassing training on the real data alone. This establishes RealClass as a valuable companion resource to mitigate data scarcity.

% We also note limitations in the current pipeline, particularly the interaction between synthetic noise and reverberation, which may result in mismatched acoustic fields. Addressing this through more consistent simulation will be the focus of future work.

We aim to make RealClass publicly available to support the research community and enable further advances in robust classroom speech technologies.

\bibliographystyle{IEEEbib}
\bibliography{strings,refs}

\begin{thebibliography}{10}

\bibitem{radford2023robust}
Alec Radford, Jong~Wook Kim, Tao Xu, Greg Brockman, Christine McLeavey, and Ilya Sutskever,
\newblock ``Robust speech recognition via large-scale weak supervision,''
\newblock in {\em International Conference on Machine Learning}. PMLR, 2023, pp. 28492--28518.

\bibitem{san2024predicting}
Nay San, Georgios Paraskevopoulos, Aryaman Arora, Xiluo He, Prabhjot Kaur, Oliver Adams, and Dan Jurafsky,
\newblock ``Predicting positive transfer for improved low-resource speech recognition using acoustic pseudo-tokens,''
\newblock {\em arXiv preprint arXiv:2402.02302}, 2024.

\bibitem{nowakowski2023adapting}
Karol Nowakowski, Michal Ptaszynski, Kyoko Murasaki, and Jagna Nieuwa{\.z}ny,
\newblock ``Adapting multilingual speech representation model for a new, underresourced language through multilingual fine-tuning and continued pretraining,''
\newblock {\em Information Processing \& Management}, vol. 60, no. 2, pp. 103148, 2023.

\bibitem{coppa2024}
{Federal Trade Commission},
\newblock ``Children's online privacy protection rule (coppa),'' 1998,
\newblock Accessed: 2024-10-12.

\bibitem{attia2024cpt}
Ahmed~Adel Attia, Dorottya Demszky, Tolulope Ogunremi, Jing Liu, and Carol Espy-Wilson,
\newblock ``Cpt-boosted wav2vec2. 0: Towards noise robust speech recognition for classroom environments,''
\newblock {\em arXiv preprint arXiv:2409.14494}, 2024.

\bibitem{southwell2024automatic}
Rosy Southwell, Wayne Ward, Viet~Anh Trinh, Charis Clevenger, Clay Clevenger, Emily Watts, Jason Reitman, Sidney D’Mello, and Jacob Whitehill,
\newblock ``Automatic speech recognition tuned for child speech in the classroom,''
\newblock in {\em ICASSP 2024-2024 IEEE International Conference on Acoustics, Speech and Signal Processing (ICASSP)}. IEEE, 2024, pp. 12291--12295.

\bibitem{pradhan2023my}
Sameer~S Pradhan, Ronald~A Cole, and Wayne~H Ward,
\newblock ``My science tutor (myst)--a large corpus of children's conversational speech,''
\newblock {\em arXiv preprint arXiv:2309.13347}, 2023.

\bibitem{shobaki2000ogi}
Khaldoun Shobaki, John-Paul Hosom, and Ronald Cole,
\newblock ``The ogi kids’ speech corpus and recognizers,''
\newblock in {\em Proc. of ICSLP}. Citeseer, 2000, pp. 564--567.

\bibitem{southwell2022challenges}
Rosy Southwell, Samuel Pugh, E~Margaret Perkoff, Charis Clevenger, Jeffrey~B Bush, Rachel Lieber, Wayne Ward, Peter Foltz, and Sidney D'Mello,
\newblock ``Challenges and feasibility of automatic speech recognition for modeling student collaborative discourse in classrooms.,''
\newblock {\em International Educational Data Mining Society}, 2022.

\bibitem{fan2024benchmarking}
Ruchao Fan, Natarajan~Balaji Shankar, and Abeer Alwan,
\newblock ``Benchmarking children's asr with supervised and self-supervised speech foundation models,''
\newblock {\em arXiv preprint arXiv:2406.10507}, 2024.

\bibitem{attia2023kid}
Ahmed~Adel Attia, Jing Liu, Wei Ai, Dorottya Demszky, and Carol Espy-Wilson,
\newblock ``Kid-whisper: Towards bridging the performance gap in automatic speech recognition for children vs. adults,''
\newblock {\em arXiv preprint arXiv:2309.07927}, 2023.

\bibitem{snyder2015musan}
David Snyder, Guoguo Chen, and Daniel Povey,
\newblock ``Musan: A music, speech, and noise corpus,''
\newblock {\em arXiv preprint arXiv:1510.08484}, 2015.

\bibitem{font2013freesound}
Frederic Font, Gerard Roma, and Xavier Serra,
\newblock ``Freesound technical demo,''
\newblock in {\em Proceedings of the 21st ACM international conference on Multimedia}, 2013, pp. 411--412.

\bibitem{cc-by-nc-sa}
{Creative Commons},
\newblock ``{Attribution-NonCommercial-ShareAlike 4.0 International (CC BY-NC-SA 4.0)},'' 2019.

\bibitem{MITOCW_YouTube}
{MIT OpenCourseWare},
\newblock ``{MIT OpenCourseWare - YouTube Channel},'' 2025,
\newblock Accessed: 2025-02-09.

\bibitem{demszky2022ncte}
Dorottya Demszky and Heather Hill,
\newblock ``The ncte transcripts: A dataset of elementary math classroom transcripts,''
\newblock {\em arXiv preprint arXiv:2211.11772}, 2022.

\bibitem{reimers-2019-sentence-bert}
Nils Reimers and Iryna Gurevych,
\newblock ``Sentence-bert: Sentence embeddings using siamese bert-networks,''
\newblock in {\em Proceedings of the 2019 Conference on Empirical Methods in Natural Language Processing}, 2019.

\bibitem{douze2024faiss}
Matthijs Douze, Alexandr Guzhva, Chengqi Deng, Jeff Johnson, Gergely Szilvasy, Pierre-Emmanuel Mazaré, Maria Lomeli, Lucas Hosseini, and Hervé Jégou,
\newblock ``The faiss library,''
\newblock 2024.

\bibitem{SteamAudio}
{Valve Corporation},
\newblock ``{Steam Audio: Real-time Physics-Based Spatial Audio},'' 2025,
\newblock Accessed: 2025-02-09.

\bibitem{nieminen2021unity}
Topi Nieminen,
\newblock ``Unity game engine in visualization, simulation and modelling,''
\newblock {B.S.} thesis, 2021.

\bibitem{farina2000simultaneous}
Angelo Farina et~al.,
\newblock ``Simultaneous measurement of impulse response and distortion with a swept-sine technique,''
\newblock {\em Preprints-Audio Engineering Society}, 2000.

\bibitem{hsu2021robust}
Wei-Ning Hsu, Anuroop Sriram, Alexei Baevski, Tatiana Likhomanenko, Qiantong Xu, Vineel Pratap, Jacob Kahn, Ann Lee, Ronan Collobert, Gabriel Synnaeve, et~al.,
\newblock ``Robust wav2vec 2.0: Analyzing domain shift in self-supervised pre-training,''
\newblock {\em arXiv preprint arXiv:2104.01027}, 2021.

\end{thebibliography}

\end{document}